\documentstyle[multicol,psfig,aps]{revtex}
\begin{document}
\draft

\title{$\beta^+$ and electron capture decay of  fp shell nuclei with $Tz=-2$}
\author{{\bf E. Caurier, F. Didierjean 
\footnote[1]{Present address: \\
                              Eurisys Mesures,
                              67380 Lingolsheim, France.},
         F. Nowacki 
\footnote[2]{Present address: \\ 
                           Laboratoire de Physique Th\'eorique,
                           3-5 rue de l'Universit\'e, 67084 Strasbourg-Cedex, France.}
        and G. Walter}\\
      IRES IN2P3-CNRS/Universit\'e Louis Pasteur BP20,\\ 
      F-67037 Strasbourg-Cedex, France.\\
}
\date{\today}
\maketitle 

\begin{abstract} 
Allowed $\beta^+$ branches of very proton-rich $fp$ shell $Tz=-2$ nuclei 
at the proton drip-line are calculated in the full fp valence space.
The $\beta^+$ decay half-lives calculated with the standard quenching
factor ($g^{eff}_{A}/g_{A}$)=0.74 are in good agreement with
existing experimental data. Detailed branching Gamow-Teller
strength are predicted but comparison with experiment  is still difficult since,
in most cases, spectroscopic information is not yet available.
 
\end{abstract} 

\begin{multicols}{2}
\section{Introduction}
  
The continuous development of experimental techniques and in particular 
those of Radioactive Ion Beam facilities allow now studies at the drip lines.
Since Coulomb interaction strongly limits spatial extension of 
proton rich nuclei, the limit of stability 
on this side of the nuclear chart is closer
to accessibility. Moreover, going on the proton rich side, $Q_{\beta^+}$
energy window increases, giving access to large Gamow-Teller
 strength. This latter is
a very sensitive test of the microscopic structure of many-body wave
functions, but also of the different correlations which can reduce the 
observable strength.

In the $sd$ shell, Muto et al. \cite{Mu:91} have studied
the $\beta^+$ decay of several proton rich nuclei. With
the Wildenthal interaction \cite{Wi:85}, they nicely reproduced
half lives for nuclei in the range of $T_{z}=-1$ to $T_{z}=-3$.
A global quenching factor of 0.60 was used for GT transitions. Nevertheless,
deviation of this overall value was observed by direct comparison
of strength distribution up to the giant resonance region. This behaviour,
also observed in the $fp$ shell was recently analysed  \cite{que:95}
as the influence at high energy of intruder strength mixing.

Ormand \cite{Orm:96} has also recently 
investigated properties of proton drip-line
nuclei, at $sd-fp$ interface. Calculated $\beta$ decay half lives   
were in overall good agreement with available experimental ones.
Predicted branching ratios for $\beta$ decay were given.
Particular attention  was made to determine the best candidates for the
observation of correlated two-proton emission. 
These turned to be $^{38}Ti$ and $^{45}Fe$.
Quite recently, these calculations have been extended by Cole \cite{Col:96}
to both heavier and lighter nuclei. As a result, $^{34}Ca$, $^{48}Ni$
and  $^{54}Zn$ have also be proposed as nuclei unstable to diproton decay.

Theoretical studies give a microscopic description and a successful
account of many observables for stable or unstable $fp$ shell nuclei.
In particular, detailed shell model calculation of the GT strength for 
$A=48, \; 47-49, \; 54 \text{ and } 56$ \cite{a48:94,a47:97,fe54:95}
have been made and a good agreement has been found with the available
(p,n) data.      
In the line of these studies, 
we investigate now the GT strength through the weak probe,
looking into the beta decay properties of $ Tz=-2$
nuclei from $^{44}Cr$ to $^{56}Zn$ (see Figure 1). As $^{54}Cu$ is predicted to be 
unstable, it was not included in our study.

Recent significant advances have been made in the spectroscopy of
these very proton-rich nuclei \cite{Bor:92,Bor:92b,Fau:95,Fau:96}.
But if half-lives could
be measured in all cases, except  $^{56}Zn$, the branching strength
are still poorly known. With the  available data, the analysis of the 
proton spectra reveals the decay of the IAS, following the Fermi beta transition,
with in addition one or two peaks which describe only partly the decay
of the states populated through Gamow-Teller transitions. Information
on gamma transitions
subsequent to, or in competition with the proton emission
of the beta populated states, is not yet available.

\section{Valence Space, operators and procedure} 
 
We consider the $fp$ shell for protons and neutrons.
All the calculations are done in the full
space up to the case of the decay $^{48}Fe$.
For $A=50,52 \text{ and }56$, we  calculate the 
ground state and the Gamow Teller sum rule in the full space, but  we need 
to introduce a truncation
scheme for the study of the detailled strength function.
 With the usual notations where $f$ stands for  $f_{7/2}$,
r for any of the other subshells ($p_{1/2}$,$p_{3/2}$,$f_{5/2})$ and t
for the number of particle excitations in the upper shells, we
restrict ourselves  to  configurations ``at the level t''
which includes configurations :   
\begin{eqnarray*}
 f^{n-n_0}r^{n_0}+f^{n-n_0-1}r^{n_0+1}+...+f^{n-n_0-t}r^{n_0+t}.
\end{eqnarray*}
At the level of truncation ``t'', we keep in the space only configurations
with a maximum of t particles promoted from $f_{7/2}$ to
$p_{1/2}$,$p_{3/2}$,$f_{5/2}$ shells.
We can remark that the ${\bf \sigma}$ operator which connects spin-orbit
partners ($f_{7/2}$ and $f_{5/2}$) introduces configurations at the level t+1.
 Moreover, $T_{+}$ operator for Z$>$28 also
introduces configurations  at the level t+1.

We use the KB3 interaction
 \cite{Pov:81}, a minimally modified version of Kuo-Brown interaction.
This latter interaction was cured to fix correct single particle
energies around $^{48}Ca$ and $^{56}Ni$. As already
mentionned, it gives an excellent agreement with spectroscopic data
in the region of stable nuclei. 
Shell model matrices are built and diagonalized with the code ANTOINE (m-scheme)
\cite{Cau:90} and a new shell model code operating in coupled
scheme representation \cite{Now:95,Nowb:95}.
Gamow-Teller operators are renormalized by the standard value in the
$fp$ shell : $(0.736)^2$ corresponding to the systematics of \cite{Mar:96}
and taking into account the recent measurement of ratio of
axial over vector coupling constants 
\cite{Abe:97},

\begin{eqnarray*}
& &B(GT)=\left( \frac{g_A}{g_V} \right)_{eff}^2 \times 
\langle \sigma \tau \rangle^2 \\
& & \text{ with} 
\left( \frac{g_A}{g_V} \right)_{eff}=(0.736)\times 
\left( \frac{g_A}{g_V} \right)_{free} \\
& & \text{and} 
\left( \frac{g_A}{g_V} \right)_{free}=-1.274(3). \\
\end{eqnarray*}

The half-lives are given by :
\begin{eqnarray*}
& &(f^A+f^{\epsilon})t= \frac{6146 \pm 6}{\left(\frac{f_V}{f_A} \right)
B(F)+B(GT)},\\
& &\text{ with here } B(F)= \langle  \tau \rangle^2 = 
\left[ T(T+1) - T_{zi}T_{zf } \right]
\end{eqnarray*}

The ground state $Q_{EC}$ values are obtained with the atomic masses
taken from the compilation by Audi and Wapstra \cite{Aud:95}. 
For these nuclei, far from stability, reported values (and errors)
result from systematic trends. For the evaluation of the Fermi
transitions, the energy
difference between pairs of isobaric analog levels is obtained with
the Coulomb displacement energies recently tabulated by Antony et al.
\cite{Ant:97}. In this work, for levels with isobaric spin $T=2$, 
fits to the most recent data yield to the equation:
\begin{eqnarray*}
& & \Delta E_{C}=1406.7(6)\frac{Z}{A^{\frac{1}{3}}}-872.8(32)  \\
\end{eqnarray*}
where $Z$ is the average charge of the pair and  A the mass number.

\section{B(GT) results}

Gamow-Teller sum rule $S^+$, calculated in the full space are given in table[1].
It can be compared to $S^-$ in the mirror nucleus 
given by charge exchange reactions $(p,n)$  \cite{Wan:88,Rap:83} which
make possible to observe , in principle, the total Gamow-Teller
strength distribution .

The increase of $S^+$ with the mass number, reported in Table 1,
corresponds to the expected variation in this part of the $fp$ shell
as the number of valence protons increases with A and the number of
neutrons holes is high. If we plot the calculated $S^+$ values 
versus the product of protons, $Z_{val}=Z-20$ and the number of
neutron holes in the full $fp$ shell, $40-N$ (see Figure 2), the value
increases, roughly linearly, as shown already by Koonin and Langanke \cite{Koo:94}
who discussed the experimental results on $S^+$ in mid-$fp$ shell nuclei. 

Strength functions are obtained with 
the Whitehead method \cite{Whi:80}
by a Lanczos procedure on an initial state taken as the sum rule state
$|\Sigma \rangle = \sigma \tau_+ |\psi_{initial} \rangle$.  
This produces the splitting  of the sum rule state over physical 
states in the daugther nucleus.
To approximate Gamow-Teller strength distribution with enougth accuracy,
we need to perform a large number of Lanczos iterations.
For each decay of the calculation, we have performed 300 iterations, which
allow a correct convergence of all low-lying eigenstates taking part
into the decay.
Due to this procedure and to large dimensionalities in $A=50,52 \text{ and }56$,
 strength functions 
were performed in a restricted configurations space corresponding to $t=4$.

In Figure 3, the $B(GT)$ distribution is shown. It reveals a very high density
of transitions, especially  in the case of the decay of the odd-odd nuclei 
$^{46}Mn \text{ and }
^{50}Co$, were the initial non zero angular momentum is connected to three
different  final values.
A resonant-like structure is appearing with a maximum below 
the $Q_{EC}$ limit.
It should be noticed that most of these daugther states
are proton unbound and that the observation should be reached
by beta-delayed proton delayed spectroscopy. Gamma detection should
also be 
associated to take into account the radiative width of the levels
populated in these decays. 

\section{Beta Decay lifetimes}

Half-lives were obtained from the partial lives
of each GT decay and from Fermi decay partial life.
Figure 4 details the beta branching ratio of each decay i.e. 
the relative contribution of each decay to the total half-life.
Attention was paid to reach convergence for each contribution 
in the cases of even-even parent decay.
An arbitrarily cut off on the vertical scale was set at $10^{-4}$
and could represent the experimental sensitivity.
 
If we consider the general features of the decays we have studied, we
note first the significant contribution of the Fermi transition in the
decay rate, in particular for the even-even nuclei. We also note (see
fig. 4) the strong decrease of the beta-branch intensity with the
excitation energy in the daughter nucleus. This fall introduces an
experimental limit, far below the $Q_{EC}$ value.

The theoretical calculation and experimental values are compared
in table [2]. Except for $^{56}Zn$, half-lives have been 
measured and are found in  good agreement with shell model
estimates, in particular for even-even nuclei.

For comparison, we have also reported two different predictions:
the results of the gross theory of beta decay of 
Tachibana et al. \cite{Tac:88} for a given set of
their parametrization  and also the recent
evaluation of the half-live with respect to Gamow-Teller $\beta$
decay calculated by P. M\"oller et al. \cite{Mol:97} from 
a Quasi-particule Random Phase Approximation. The QRPA estimates give a good agreement for
$^{48}Fe$ and $^{50}Co$ but large deviations are found in the other decays.
Tashibana model is of macroscopic nature involving
a set of dedicated parameters and gives in this mass region a satisfactory
description of  the decay lifetimes.

In most  cases of proton rich nuclei, the large Q value allows the daugther nucleus 
to decay mainly via proton emission. 
Nevertheless,  
in the particular case of the IAS of even-even parent, we have computed M1 transitions and
gamma widths. These results
are shown is table III. The strongest transition between $0^+(T=2)$
and $1^+(T=1)$ is only reported here
and the gamma energy is taken from the mirror nucleus. It is found that the calculated
strength are close to the upper limit which has been observed for $M1$
isovector transitions \cite{end:93}.
In previous experimental work \cite{Bor:92,Fau:96,Fau:94},
a deficit in the intensity of the emitted protons from the IAS
was reported for the case of $^{52}Ni$. Gamma dexcitation are competing
with particle emission and an accurate determination of 
isospin violation in the proton decay with INC interaction as 
performed previously by by Ormand \cite{Orm:96}
would allow to determine the relative intensity of the two processes.

\section{Conclusion} 

The strength function, branching ratios and half-life of the $T_z=-2$
nuclei  ($44 \leq A \leq 56$) have been computed in a full major
oscillator shell. The calculation gives a prediction of
the distribution of the Gamow-Teller strength with a maximum below,
or very close to the $Q_{EC}$ window limit. The experimental data
available do not allow to test the calculated distributions but, only
to compare half-lifes. The 
comparison with other calculations like models based on the finite-range
droplet model made for a very large range of nuclei, is of interest to
test the reliability of the calculations for nuclei far from
stability, and their sensitivity to the details of nuclear
structure.\\ 
It is important to note that the same shell model calculations
give an account of the Gamow-Teller strength observed in
the $(p,n)$ reactions on $T_z=+2$ targets and the one resulting from
the beta decay of $T_z=-2$ nuclei.
The good agreement between the calculated values and experimental ones,
observed in the $(p,n)$ data of $A=56$, cannot be tested up to now
for the beta decay.\\
We conclude with a few  remarks concerning the problems related to the
experimental study of these very proton-rich nuclei. First, the
calculations indicate a strong splitting of the strength which makes
more difficult the observation of the corresponding branches and impose
severe limits for the background. The second remark concerns the range
of the detectable beta branches which is more limited than the
$Q_{EC}$ window and makes essentially unobservable transitions in the
last fraction of this window. Only a dramatic improvement in the
production rates of these very proton-rich nuclei would modify this
limitation and make possible the observation of the shape of the main
components of the Gamow-Teller distribution by beta-decay study. Finally,
the calculation reveals the strength of the $M1$ transitions
crresponding to the radiative decay of the IAS populated by beta decay
and the neccessity of gamma detection for a complete experimental survey.

\end{multicols}

\newpage

 \begin{table}[h]
 \begin{center} 
 \begin{tabular}{lcccccc}\hline\hline \\[-5pt] 
 nucleus& $^{44}$Cr & $^{46}$Mn & $^{48}$Fe & $^{50}$Co & $^{52}$Ni& $^{56}$Zn \\ [5pt]
\hline \\[-5pt]
$S^+=\sum_{i} GT^+_i$    & 12    & 12.89 & 13.26 & 14.68 & 15.33 & 16.69 \\ [5pt]
$S^+_{renorm.}$    & 6.50  & 6.98 & 7.18 & 7.95 &8.30 & 9.04 \\ [5pt]
$S^+_{exp.} (\text{ from } S^-_{exp.}(p,n))$ &    &  &  &  & 5.9 $\pm$ 1.5$^a$ & 9.9 $\pm$ 2.4$^b$ \\ [5pt]
\hline \hline
 \end{tabular}
 \vspace*{0.5cm}
 \caption[]{Total value of the calculated $\beta^+$
GT transition strength for $T_z=-2$ $fp$-shell nuclei
and renormalized value obtained with the standard quenching factor.\\
$^a$ Reference \cite{Wan:88} \\
$^b$ Reference \cite{Rap:83} \\
}
 \end{center}
 \end{table}
 
 \begin{table}[h]
 \begin{center} 
 \begin{tabular}{ccccccc}\hline\hline \\[-5pt] 
 nucleus& $^{44}$Cr & $^{46}$Mn & $^{48}$Fe & $^{50}$Co & $^{52}$Ni& $^{56}$Zn \\ [5pt]
\hline \\[-5pt]
$t_{1/2}^{fermi}$ (ms)  & 132   & 104   &   98  &   83  &  76   &  56   \\ [5pt] 
$t_{1/2}^{GT} $(ms)     & 83    &  39   &  114  &   40  &  146  &  42   \\ [5pt] 
$T_{1/2}^{tot}$ (ms)    & 51    &  29   &   53  &   27  &  50   &  24  \\ [5pt] 
$T_{1/2}^{exp}$ (ms)    & 53$^{+4 a}_{-3}$  & 41$^{+7 a}_{-6}$ & 44$\pm$7 $^b$  &
44$\pm$4 $^b$ & 38$\pm$5 $^c$ &-\\ [5pt] 
&&&  &&&\\ [5pt]
$t_{1/2}^{GT}$ (ms)     & 119   &  15    &  60  &  47   &  77   &  83  \\   [5pt] 
(M\"oller)$^d$&&&&&&\\ [5pt]  
$T_{1/2}$ (ms)  & 83    &  53    &  48  &  36   &  35   &  24   \\  [5pt]
(Tashibana)$^e$    &       &        &      &       &       &       \\ [5pt] 
\hline \hline
 \end{tabular}
 \vspace*{0.5cm}
 \caption[]{The Fermi and
GT components of the calculated half-lives are given with the global value.
The available experimental values are also presented as well as, in the
last two lines, the predicted values from the QRPA model and the gross theory.\\
$^a$ Reference \cite{Bor:92} \\
$^b$ Reference \cite{Fau:96} \\
$^c$ Reference \cite{Fau:94} \\
$^d$ Reference \cite{Mol:97} \\
$^e$ Reference \cite{Tac:88} \\
}
 \end{center}
 \end{table}

 \begin{table}[h]
 \begin{center} 
 \begin{tabular}{ccccccc}\hline\hline \\[-5pt] 
 nucleus& $^{44}$V & $^{46}$Cr & $^{48}$Mn & $^{50}$Fe & $^{52}$Co& $^{56}$Cu \\ [5pt]
\hline \\[-5pt]
$E_{\gamma}$ (MeV)      & 2.12   & - & 2.57 & - & 2.02 & 1.81 \\ [5pt]
$B(M1) (\mu^2_N)$  & 5.57   & - & 2.87 & - & 1.87 & 1.58 \\ [5pt]
$\Gamma$ (eV)      & 0.627  & - & 0.58 & - & 0.18 & 0.11 \\ [5pt]
$\tau_m$ (fs)     & 1.06   & - & 1.15 & - & 3.63 & 5.99 \\ [5pt]
$\Gamma/\Gamma_m$ (W.u.)&3.12& - & 1.40 & - & 1.04 & 0.88 \\ [5pt]
\hline \hline
 \end{tabular}
 \vspace*{0.5cm}
 \caption[]{Radiative decay calculated in the daugther of the even-even emitter.
 Calculated transitions strength are also reported (in $ \mu^2_N$ and W.u.)
 with corresponding width and meanlife values.} 
\end{center}
 \end{table}

\newpage

\begin{figure*}[here]
\psfig{file=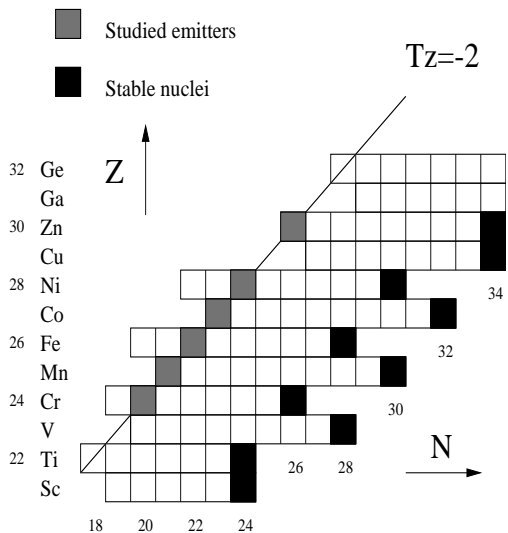,height=7cm ,width=8cm}
\caption[]{Proton drip-line and studied nuclei}
\end{figure*}

\begin{figure*}[here]
\psfig{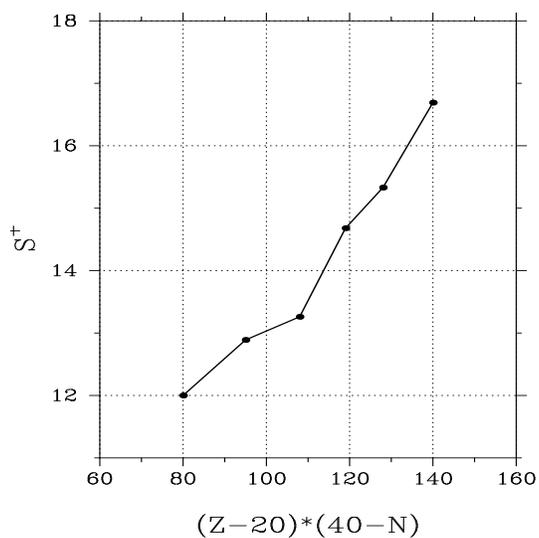}
\caption[]{$S^+$ as function of number of particles
and holes in the valence space.}
\end{figure*}

 \newpage
  
  \begin{figure*}[top]
   \caption[t]{Theoretical Gamow-Teller strength in units of GT sum rule (x-axis  in MeV). }
   \psfig{file=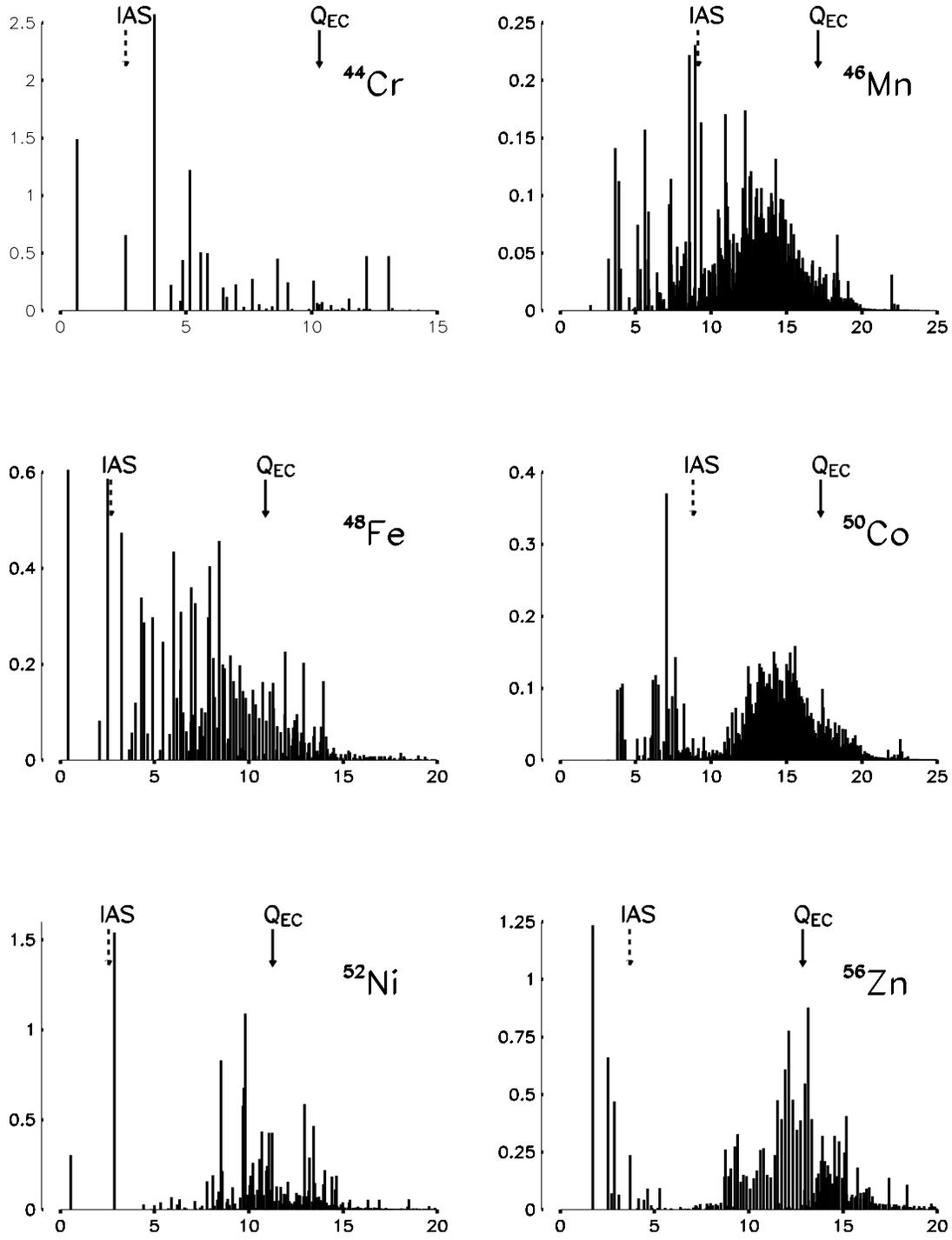,height=21cm ,width=22cm}
   \end{figure*}
   
 \newpage
 
   \begin{figure*}[top]
   \caption[t]{Theoretical branching ratio (x-axis  in MeV, note the change of scale
                                            with FIG. 3).}
   \psfig{file=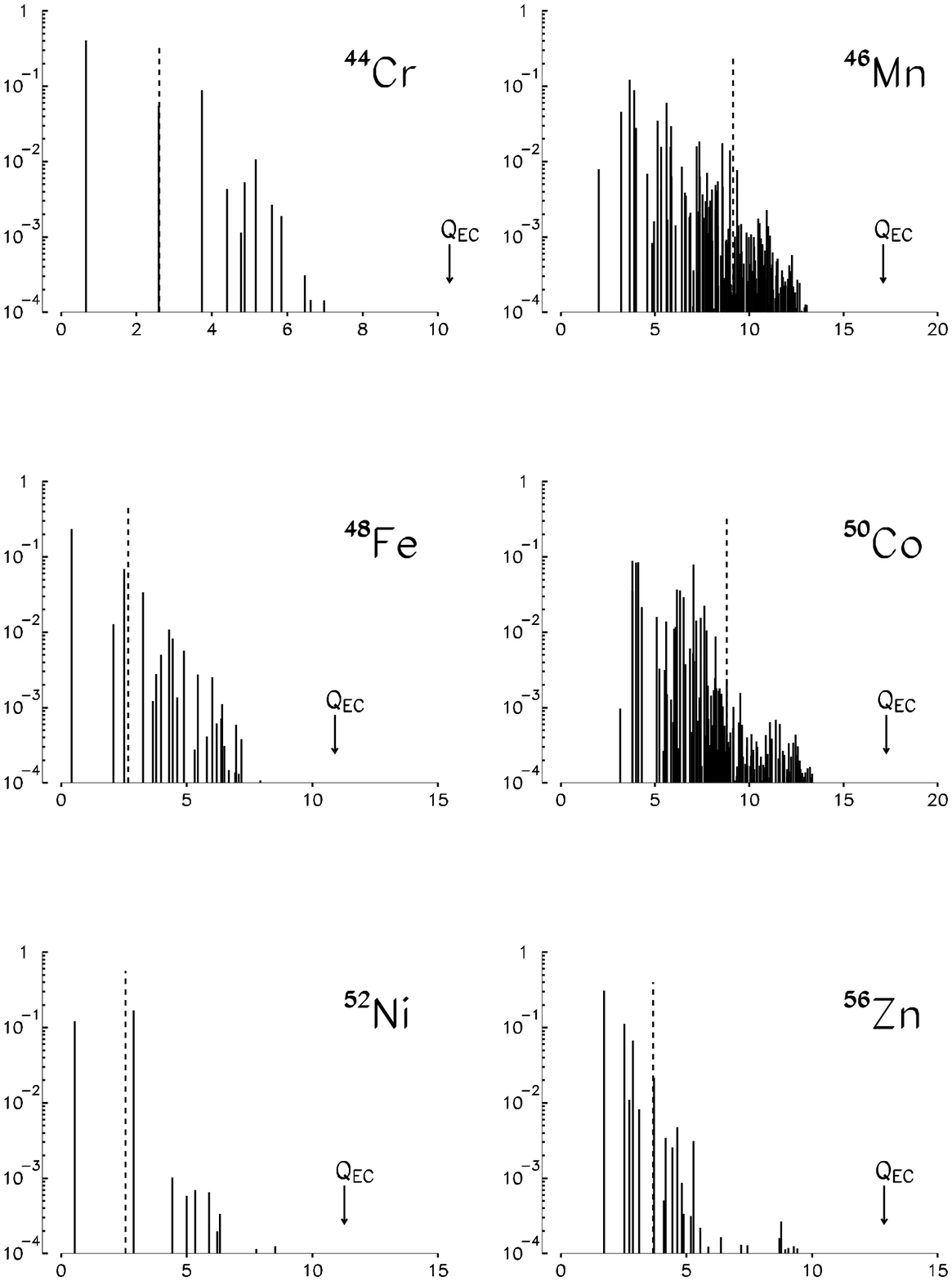,height=21cm ,width=22cm}
   \end{figure*}

\begin{thebibliography}{99} 
\bibitem{Mu:91}  K. Muto, E. Bender and T. Oda
                 Phys. Rev. C{\bf 43} (1991)1487.
\bibitem{Wi:85} B. A. Brown and B. H. Wildenthal, At. Data Nucl. Data
                Tables {\bf 33}(1985)347.
\bibitem{que:95} E. Caurier, A. Poves and A. P Zuker
                 Phys. Rev. Lett. {\bf 52} 1736 (1995). 
\bibitem{Orm:96}  W. E. Ormand  Phys. Rev. C{\bf 53}, 214 (1996).
\bibitem{Col:96} B. J. Cole, Phys. Rev. C {\bf 54}, 1240 (1996).
\bibitem{Pov:81} A. Poves and A. Zuker, Phys. Rep. 70, 4 (1981). 
\bibitem{Mar:96} G. Martinez-Pinedo, A. Poves, E. Caurier and A. P Zuker,\\
                 Phys. Rev. C{\bf 53}, R2602 (1996). 
\bibitem{Aud:95}   G. Audi and A. H. Wapstra, Nucl. Phys. A{\bf 595}, 409  (1995)
\bibitem{Ant:97} M. S. Antony, A. Pape and J. Britz,\\ 
                 At. Data Nucl. Data Tables {\bf 66}, 131 (1997).
\bibitem{Wan:88} D. Wang, J. Rapaport, D. J. Horen, B. A. Brown, C. Gaarde,\\
                 C. D. Goodman, E. Sugarbaker, T. N. Taddeucci, \\
                 Nuc. Phys. {\bf A480}, 285 (1988).
\bibitem{Rap:83} Rapaport et al., Nuc. Phys. {\bf A410}, 371 (1983).         
\bibitem{Koo:94}  S. E. Koonin and K. Langanke, Phys. Lett. B{\bf 326}, 5 (1994). 

\bibitem{Cau:90} E. Caurier, code ANTOINE (m scheme), Strasbourg (1989).
\bibitem{Now:95} F. Nowacki, E. Caurier, code (coupled scheme), Strasbourg (1995).
\bibitem{Nowb:95} F. Nowacki, Ph. D. Thesis, (Strasbourg,1995).
\bibitem{a48:94} E. Caurier, A. P Zuker, A. Poves and G. Mart\'{i}nez-Pinedo,
                 Phys. Rev. C{\bf 50}, 225 (1994). 
\bibitem{a47:97} G. Mart\'{i}nez-Pinedo, A. P Zuker, A. Poves and E. Caurier,\\
                 Phys. Rev. C{\bf 55}, 187 (1997). 
\bibitem{fe54:95} E. Caurier, G. Mart\'{i}nez-Pinedo, A. Poves and A. P Zuker,\\
                 Phys. Rev. C{\bf 52}, 1736 (1995). 

\bibitem{Fau:94} L. Faux et al., Phys. Rev. C{\bf 49}, 2440 (1994).   
\bibitem{Bor:92} V. Borrel {\em et al.}, Z. Phys. A{\bf 344} (1992)135.    
\bibitem{Bor:92b} V. Borrel {\em et al.}, Proc. 6th Int. Conf. on Nuclei
                 Far From Stability and 9th Int. Conf. on Atomic Masses
                 and Fundamental Constants, Bernkastel-Kues, 1992;
                 Inst. Phys. Conf. Ser. No 132: section3.  

\bibitem{Fau:95} L. Faux , Ph. D. Thesis (Bordeaux, 1995). 
\bibitem{Fau:96} L. Faux et al., Nucl. Phys. {\bf A602}, 167(1996).  

\bibitem{Whi:80} R. R. Whitehead, in Moments Methods in Many Fermion 
                 Systems,\\
                 edited by B. J. Dalton, S. M. Grimes, J. D. Vary 
                 and S. A. Williams \\
                 (Plenum, New York, 1980), p. 235.
\bibitem{Abe:97} H. Abele et al., Phys. Lett. B407, 212 (1997).   
\bibitem{Tac:88} T. Tachibana, M. Yamada and K. Nakata,\\
                 Report of Sci. and Res. Lab., \\
                 Waseda University 88-3 (1988).
\bibitem{Mol:97} P. M\"oller, J. R. Nix, and K.-L. Kratz, \\
                 At. Nucl. Data Tables {\bf 66}, 131 (1997).

\bibitem{end:93} P. M. Endt,
                 At. Nucl. Data Tables {\bf 55}, 171 (1993).

\end{thebibliography}
\end{document}